# Human Behavior Recognition Method Based on CEEMD-ES Radar Selection

Zhaolin Zhang
*School of electronics and information engineering*
*Harbin institute of technology*
Harbin, China
aeryszzl@163.com

Mingqi Song
*School of information science and engineering*
*Harbin institute of technology (Weihai)*
Weihai, China
songmq95@163.com

Wugang Meng
*Collage of computing*
*Georgia institute of technology*
Atlanta, U.S.A
https://orcid.org/0000-0003-4289-2409

Yuhan Liu
*School of information science and engineering*
*Harbin institute of technology (Weihai)*
Weihai, China
yuhanliu1210@163.com

Fengcong Li
*School of electronics and information engineering*
*Harbin institute of technology*
Harbin, China
xialulee@sina.com

Xiang Feng
*School of electronics and information engineering*
*Harbin institute of technology*
Harbin, China
fengxiang23031@163.com

Yinan Zhao
*School of electronics and information engineering*
*Harbin institute of technology*
Harbin, China
zhaoyinan@hit.edu.cn

*Abstract*—In recent years, the millimeter-wave radar to identify human behavior has been widely used in medical, security, and other fields. When multiple radars are performing detection tasks, the validity of the features contained in each radar is difficult to guarantee. In addition, processing multiple radar data also requires a lot of time and computational cost. The Complementary Ensemble Empirical Mode Decomposition-Energy Slice (CEEMD-ES) multistatic radar selection method is proposed to solve these problems. First, this method decomposes and reconstructs the radar signal according to the difference in the reflected echo frequency between the limbs and the trunk of the human body. Then, the radar is selected according to the difference between the ratio of echo energy of limbs and trunk and the theoretical value. The time domain, frequency domain and various entropy features of the selected radar are extracted. Finally, the Extreme Learning Machine (ELM) recognition model of the ReLu core is established. Experiments show that this method can effectively select the radar, and the recognition rate of three kinds of human actions is 98.53%.

## I. Introduction

Recent developments in the field of semiconductor technology have led to a renewed interest in civil radar what is gradually moving into the miniaturization stage. Compared with other sensors, radar can effectively resist the effects of heat sources, light sources, and extreme weather. Among many radars, millimeter-wave radar also has the advantages of low power, small size, and high bandwidth. Therefore, it is widely used in detection technologies such as autonomous driving, medical care, assisted care, and human-computer interaction technologies. In private spaces, millimeter-wave radar eliminates the need for wear and enables adequate protection of personal privacy. By detecting the Doppler frequency shift of various parts of the human body, human movements and behaviors are recognized [1]-[3].

Amin et al. [4] explored the relationship between Doppler and micro-Doppler with time, the subtle feature of cane in gait time-frequency features, and verified the effectiveness of this feature in distinguishing cane gait from normal human gait.

Sakagami et al. [5] extracted features from distance-Doppler, time-Doppler, and time-distance images for motion recognition using CNN, and the results showed that multi-domain signals could improve classification accuracy. Zihao et al. [6] also used the above three images, fused the three image features by building a data cube, and used CNN for human behavior recognition, verifying that the data cube with 3D information helps to improve the recognition accuracy. Javier et al. [7] used linear predictive coding (LPC) to extract different frequency mixed micro-Doppler features, which reduces the time frames required to acquire the features and reduces the time cost. However, the above studies were only in the radar frontal illumination scenario and did not consider the effect of the human body on the radar echo when it was in different positions. Li et al. [8] detected the human body by UWB pulsed radar in three different positions, extracted the features in the time-frequency map and cepstrum map, and used Bi-directional Long Short-Term Memory (Bi- LSTM) for human behavior recognition. Craley et al. [9] utilized two radars for human detection and extracted the joint time-frequency map as features for recognition by LSTM. Fioranelli et al. [10-11] set up multiple radar nodes and used two different fusion methods, feature fusion and decision-level fusion, to achieve higher recognition accuracy than a single radar. The importance of radar selection under special conditions were mentioned in the prospects. There are many benefits to using multiple radars, but they can still cause problems such as increased computation and time costs, invalidation under special conditions, and so on.

In summary, in order to save various costs required for human behavior recognition and ensure the recognition effect. Use Complementary Ensemble Empirical Mode Decomposition (CEEMD) signal decomposition method[12]. This paper proposes a multistatic radar selection method based on CEEMD-Energy Slice (CEEMD-ES), which uses the difference in energy ratio between limbs and trunk to select radar signals with prominent human behavior characteristics. Then, extract the time domain, frequency domain, and various entropy features of the radar signal that are used to characterize human behavior. Finally, by establishing the extreme learning machine (ELM) model of the ReLu core[13], The validity of the CEEMD-ES radar selection method and the extracted entropy features is verified.

This work was supported in part by the National Natural Science Foundation of China under Grant 61771160, and the project ZR2019BF037 supported by Shandong Provincial Natural Science Foundation.

The rest of this paper is organized as follows. Section II presents the CEEMD-ES algorithm and feature extraction. Division III offers the experiments and verifies the effectiveness of the method. Finally, the paper is summarized.

## II. BASIC ALGORITHM

### A. CEEMD-ES Alorithm

The body radar echo signal is defined as :

$$x(t) = Arect\left(\frac{t}{\tau}\right)e^{j2\pi\left(f_0 t+\frac{1}{2}ut^2\right)} \quad (1)$$

where $u(t) = Arect\left(\frac{t}{\tau}\right)e^{j\pi ut^2}$ is the complex envelope of the signal, $rect\left(\frac{t}{\tau}\right)$ is the rectangular function, $\tau$ is the pulse width, $u$ is the slope of the change of the instantaneous frequency of the signal, and $f_0$ is the center frequency.

Then the calculation process for CEEMD-ES can be expressed as:

*1) $n(t)$ is defined as the Gaussian white noise signal. The mathematical expression of the divided radar signal can be defined as:*

$$x_1(t) = Arect\left(\frac{t}{\tau}\right)e^{j2\pi\left(f_0 t+\frac{1}{2}ut^2\right)} + n(t) \quad (2)$$

$$x_2(t) = Arect\left(\frac{t}{\tau}\right)e^{j2\pi\left(f_0 t+\frac{1}{2}ut^2\right)} - n(t) \quad (3)$$

*2) Extract the upper envelope $e_1^u$, $e_2^u$ of $x_1(t)$ and $x_2(t)$, the lower envelope $e_1^d$, $e_2^d$ of $x_1(t)$ and $x_2(t)$. Then the mathematical expression of their average value is given by*:

$$e_1^{avg} = e_1^u + e_1^d \quad (4)$$

$$e_2^{avg} = e_2^u + e_2^d \quad (5)$$

*3) If the difference between the upper and lower envelopes is less than the set error $\vartheta$, and the number $\kappa$ which is the extreme difference points and the zero points are less than 2. Then the mathematical expression of IMF is obtained by:*

$$IMF_1 = (e_1^{avg} + e_2^{avg})/2 \quad (6)$$

*4) Update the value of $x_2(t)$ to $x_2(t) - e_2^{avg}$, repeat 2),3),4) until the iteration stops or the set screening number is reached, then the residual can be expressed as:*

$$RES = (x_1 + x_2)/2 \quad (7)$$

*5) The mathematical expression of the signal is shown in Equation (8).*

$$x(t) = \sum_{i=1}^{n} IMF_i + \gamma_n(t) \quad (8)$$

*6) Then the energy ratio of its limbs to the trunk is defined as:*

$$\sum_{i}^{n}|STFT(IMF_i(t))|/|STFT(x(t))| \quad (9)$$

### B. Feature Extraction

The time domain characteristics of the signal include Mean, Root mean square, Square root amplitude，Absolute mean, Skewness, Kurtosis, Variance, Peak-to-peak, Volatility index, Peak Impulse indicator, Margin Index, Skewness index, Kurtosis index. The frequency domain characteristics are shown in literature [14]. The approximate entropy is calculated by constructing an $m$-dimensional vector $X(t)$ of time series $x(t)$ in order with a similarity tolerance of $r$. Where $t \in (1, N)$, and $X(t)$ is defined as:

$$X(t) = [x(t), x(t+1), \cdots, x(t+m-1)] \quad (10)$$

$$d_{X(i)-X(j)} = \max_{k=0\sim m-1}|x(i+k) - x(j+k)| \quad (11)$$

$$\Phi^m(r) = \frac{1}{N-m+1}\sum_{i=1}^{N-m+1} \ln C_i^m(r) \quad (12)$$

$$C_i^m(r) = \frac{d<sum(r)}{N-m+1} \quad (13)$$

Calculation parameter Equation (11)-(13), then the approximate entropy can be calculated by Equation (14).

$$ApEn = \lim_{N\to\infty}[\Phi^m(r) - \Phi^{m+1}(r)] \quad (14)$$

The ranking entropy calculation is given by Equation (15), where ρ is some arrangement of the radar signal $x(t)$ and $p_j(\rho)$ is the probability of ρ.

$$PE(m) = -\sum_{j=1}^{m!} p_j(\rho)\ln(p_j(\rho)) \quad (15)$$

The multiscale entropy is calculated as Equation (16), where $\epsilon$ is the time scale.

$$ME_j^{(\epsilon)} = \frac{1}{\epsilon}\sum_{i=(j-1)\epsilon+1}^{j\epsilon} x_i, 1 \leq j \leq \frac{N}{\epsilon} \quad (16)$$

### C. Radar Echo Modeling and Energy Ratio Calculation of Human Target

To simplify the model, a spheroid is usually used to model the human body. A model of a spheroid can be defined as：

$$\left(\frac{x}{a}\right)^2 + \left(\frac{y}{b}\right)^2 + \left(\frac{z}{c}\right)^2 = 1 \quad (17)$$

Then the angle of incidence and azimuth of the radar are $\theta$ and $\varphi$, respectively. Its mathematical expression are $\theta = \arctan\left(\frac{\sqrt{x^2+y^2}}{z}\right)$ and $\varphi = ractan\left(\frac{y}{x}\right)$.

The RCS approximator for ellipsoidal backscattering can be expressed as:

$$RCS = \frac{\pi a^2 b^2 c^2}{(a^2 \sin^2\theta \cos^2\varphi + b^2\sin^2\theta\sin^2\varphi + c^2\cos^2\theta)^2} \quad (18)$$

Assuming that the radar is directly in front of the human body, such as the position of the radar $(0, d, 0)$, where d > b. Its expression can be reduced to Equation (19), where μ is $\frac{\sqrt{\pi}}{b}$.

$$RCS = \mu ac \quad (19)$$

$a$ is larger than $b$ in ordinary humans. Following the ellipse area formula, the RCS should be proportional to the area of the human body. As the angle of view of the radar rotates, $a$ decreases and RCS decreases accordingly, so the RCS of the human body is the largest when facing the radar in front of it. Ignoring the effect of reduced RCS in the extremities during walking, the ratio of RCS in the extremities to RCS in the whole body should be less than or equal to its area.

III. EXPERIMENT

The radar used in the experiment is placed in the position shown in Fig. 1. One is called a 0° radar when facing a person and the other is called a 90° degree radar on the side. The radar operating frequencies are 77GHz and 79GHz, respectively. The sampling frequency is 4000kHz.

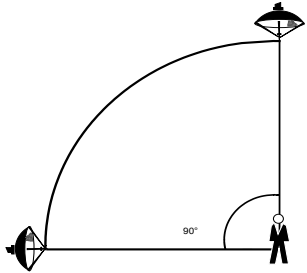

Fig. 1. Schematic diagram of radar detection position.

The measurement range of 1m. Test one person at a time. Radar is in transmit and receive mode, with a single sampling time of 4.5 seconds. Three types of motion data were collected from subjects, such as (a) single arm swing (b) swing arms alternately (c) swing legs alternately. Their time-frequency diagram are shown in Fig. 2. The periodic peak in the figure shows that the one-arm swing is cyclicity. Its trunk fluctuates less, so it is serrated at 0 frequencies and has a strong echo energy. In Fig. 2(b), both arms have a shorter period of fluctuation due to the alternating oscillation, which is characterized by continuity. In Fig. 2 (c), the alternating oscillation of the legs is similar to that of the arms, both of which are periodic. However, since the human legs account for 34% of the total skin area and the arms account for 18%[15], the RCS of the legs is considered much larger than that of the arms without considering other conditions. Therefore, the echo energy of the legs is greater than the arms. But the swing of both arms has higher stability, so the period amplitude of Fig. 2 (b) is more stable than that in Fig. 2 (c).

At the radar observation position as shown in Fig. 1, the 90° radar echo features trend is relatively small, such as in the double-arm swing whose STFT time-frequency diagram is shown in Fig. 2(d). Therefore, it is necessary to select the radar for different observation positions first.

Since the RCS of each part of the human body is different, the echo energy is also different, but because the RCS of the trunk is larger, it is challenging to select the radar based on the total echo energy. Therefore, the human trunk signal needs to be filtered. In human activities, due to the high energy of trunk echo, its time-frequency characteristics are apparent when the body is swinging wildly, and it is easy to distinguish different behavioral movements. In the area of micro-motion, researchers paid more attention to the movement state of the limbs. When the limbs are active, their frequencies tend to be greater than the trunk movement frequencies.

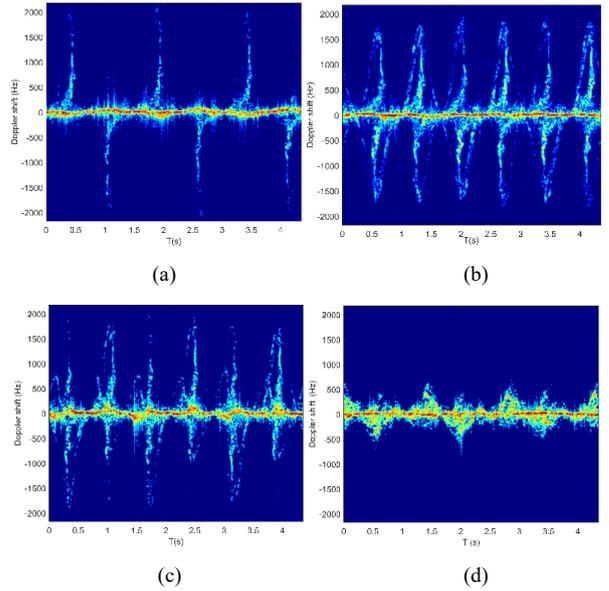

Fig. 2. Time-frequency diagram of various human behaviors. (a) Single arm swing-0°; (b) Swing arms alternately-0°; (c) Swing legs alternately-0°; (d) Swing arms alternately-90°.

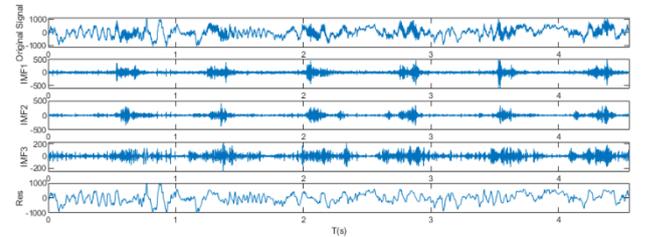

Fig. 3. Decomposition result of radar echo signal with alternating arms swinging

Take the alternating swing arm signal as an example. Firstly, selected the radar using CEEMD-ES. The IMF components were calculated by adding noise to the signal, extracting the extreme value points, solving the upper and lower envelopes. The decomposition results of the signal are shown in Fig. 3. The STFT time-frequency diagram of the Res component is shown in Fig. 4 (a). It can be seen from the figure that its low-frequency residuals are mainly the echo components of the trunk part. Therefore, it is necessary to reconstruct the other elements to find the signal of the limbs. The IMF1, IMF2, and IMF3 were summed and rebuilt, and the time-frequency diagram of the reconstructed signal is shown in Fig. 4 (b). Fig. 4 (b) has more background noise but a smaller proportion of noise energy. The Otus filtered image is shown in Fig. 4 (c) [16]. The arms fluctuate periodically. At this point, it can be considered that the trunk has been separated from the extremity signa, but because the extremities also exist in the zero frequency state, so a certain amount of extremity signal is lost.

Since the human head and trunk account for 36% of the total body area, limbs account for 64% of the total body area. Follow Equation (19), the ratio between the total energy of the limbs echo and the total energy of the echo should be about 0.64, regardless of other conditions. Its actual echo is shown in Fig. 5. As shown in illustration, the 90° radar energy ratio is slightly lower than the 0° radar. 0° radar is consistent with the actual human frontal area ratio. Therefore, 0° position radar was selected.

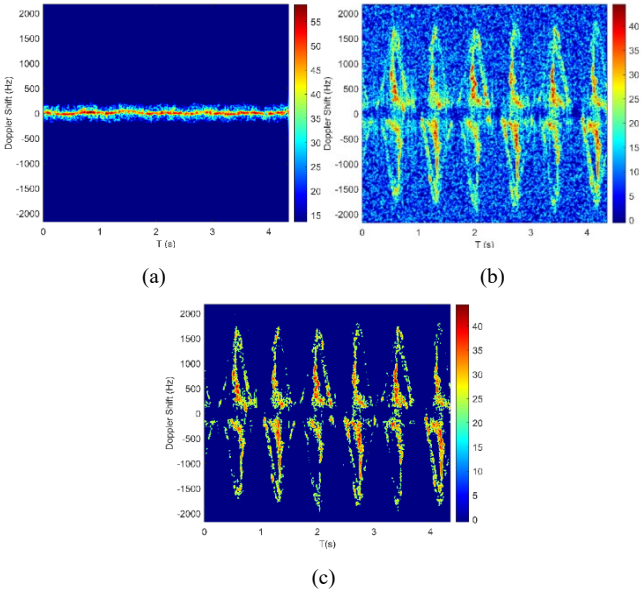

Fig. 4. Time-frequency diagram of the corresponding radar signal. (a) STFT time-frequency diagram of Res component; (b) Reconstructed signal STFT time-frequency diagram; (c) Time-frequency diagram of the reconstructed signal after Otus filtering.

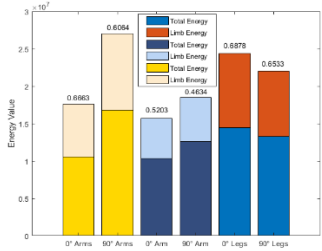

Fig. 5. Schematic diagram of limb echo and total echo energy

TABLE I. HUMAN BEHAVIOR RECOGNITION RATE

| Radar Names | Total Body Recognition Rate(%) | | | Limb (%) | Litera-ture[6] (%) |
|---|---|---|---|---|---|
| | All Features | Time-frequency Domain Features | Entropy Features | All Features | |
| 0° Radar | 98.529 | 98.215 | 98.193 | 88.267 | 96.320 |
| 90° Radar | 94.702 | 94.597 | 94.583 | 91.270 | 92.740 |

To verify the effectiveness of the proposed method, the time domain, frequency domain, and entropy features of the signal were extracted. Three behaviors were identified using ELM model of ReLu kernel. There are 120 training sets for each category and 48 test sets for each category. Repeated the experiment 1000 times to calculate the average recognition rate. The recognition rates are shown in TABLE I. . Entropy feature can improve the recognition rate of human behavior. Selecting radar by CEEMD-ES can effectively select observation radar with distinct motion characteristics and reduce time and calculation costs.

## IV. CONCLUSION

In summary, CEEMD-ES was used to separate the radar echoes from the trunk and extremities based on the significant differences in frequency between the trunk and the limbs during human exercise. Because the contribution of radar in different detection directions to recognition is different, the ratio of limb energy to total radar return energy is extracted as an index value for radar selection. The theoretical value of the index of the approximate model was calculated. By selecting a radar with a large share of limb energy, the efficiency of human behavior recognition is improved, and time and calculation costs are saved. The experimental results verify the validity of this method.

ACKNOWLEDGMENT

The authors appreciate the anonymous reviewers who help the authors to evaluate the quality of this paper.

REFERENCES

[1] Du H., Jin T., and Song Y., "DeepActivity: a micro-Doppler spectrogram-based net for human behaviour recognition in bio-radar," IET International Radar Conference, 2018, pp. 6147-6151.

[2] Liang Q, "Biologically inspired target recognition in radar sensor networks," J Wireless Com Network, pp. 1-8, 2009.

[3] M. Li, T. Chen, and H. Du, "Human Behavior Recognition Using Range-Velocity-Time Points," IEEE Access, vol. 8, pp. 37914-37925, 2020.

[4] Amin M. G., F. Ahmad, YD Zhang, and B. Boashash, "Human gait recognition with cane assistive device using quadratic time-frequency distributions," Radar Sonar & Navigation Iet, vol. 9, pp. 1224-1230, September 2015.

[5] F. Sakagami, H. Yamada, and S. Muramatsu, "Accuracy improvement of human motion recognition with MW-FMCW radar using CNN," 2020 International Symposium on Antennas and Propagation (ISAP), 2021, pp. 173-174.

[6] Y. Zhao, Z. Zhang, and Z. Zhang, "Multi-Angle data cube action recognition based on millimeter wave radar," 2020 Chinese Control And Decision Conference (CCDC), 2020, pp. 4983-4987.

[7] R. J. Javier, and Y. Kim, "Application of linear predictive coding for human activity classification based on Micro-Doppler signatures," IEEE Geoscience and Remote Sensing Letters, vol. 11, no. 10, pp. 1831-1834, Oct. 2014.

[8] H. Li, A. Mehul, J. Le Kernec, S. Z. Gurbuz, and F. Fioranelli, "Sequential human gait classification with distributed radar sensor fusion," IEEE Sensors Journal, vol. 21, No. 6, pp. 7590-7603, 15 March 15, 2021.

[9] J. Craley, T. S. Murray, D. R. Mendat, and A. G. Andreou, "Action recognition using micro-Doppler signatures and a recurrent neural network," 2017 51st Annual Conference on Information Sciences and Systems (CISS), 2017, pp. 1-5.

[10] F. Fioranelli, M. Ritchie, S. Z. Gürbüz, and H. Griffiths, "Feature diversity for optimized human Micro-Doppler classification using multistatic radar," IEEE Transactions on Aerospace and Electronic Systems, vol. 53, no. 2, pp. 640-654, April 2017.

[11] H. Li, A. Mehul, J. Le Kernec, S. Z. Gurbuz, and F. Fioranelli, "Sequential Human Gait Classification With Distributed Radar Sensor Fusion," IEEE Sensors Journal, vol. 21, no. 6, pp. 7590-7603, 15 March15, 2021.

[12] Y. Ren, P. N. Suganthan, and N. Srikanth, "A comparative study of empirical mode decomposition-based short-term wind speed forecasting methods," IEEE Transactions on Sustainable Energy, vol. 6, no. 1, pp. 236-244, Jan. 2015.

[13] A. Safaei, Q. M. J. Wu, T. Akilan, and Y. Yang, "System-on-a-Chip (SoC)-based hardware acceleration for an online sequential extreme learning machine (OS-ELM), " IEEE Transactions on Computer-Aided Design of Integrated Circuits and Systems, vol. 38, no. 11, pp. 2127-2138, Nov. 2019.

[14] J Xiao, and J Li, "Fault diagnosis model based on adaptive generalized morphological filtering and LLTSA-ELM," International Journal of Information and Communication Technology, in press.

[15] Geisheimer JL, Greneker EF, and Marshall WS., "A high-resolution Doppler model of human gait," Proceedings of the society of photo-optical instrumentation engineers, vol. 4744, pp. 8-18, 2002.

[16] N. Otsu, "A threshold selection method from gray-level histograms, " IEEE Transactions on Systems, Man, and Cybernetics, vol. 9, no. 1, pp. 62-66, Jan. 1979.